\documentstyle[prb,aps]{revtex}
\begin{document}


\draft
\title{ Enumeration of the self-avoiding polygons on \\
 a lattice by the Schwinger-Dyson equations }
\author{P. Butera\cite{pb} and M. Comi\cite{mc}}    
\address
{Istituto Nazionale di Fisica Nucleare\\   
Dipartimento di Fisica, Universit\`a di Milano\\
16 Via Celoria, 20133 Milano,  Italy} 
\maketitle
\scriptsize
\begin{abstract}    
 We show how to compute the generating function of the self-avoiding 
polygons on a lattice by using the statistical mechanics Schwinger-Dyson 
equations for the correlation 
functions of the $N$-vector spin model on that lattice.

\end{abstract} 
\normalsize

\section{Introduction}
It has long been known that   nontrivial 
 combinatorial results  concerning the enumeration of subgraphs
 are related
 to the computation of high-temperature(HT) expansions in the statistical 
mechanics of lattice spin models\cite{combi}. 

From a physicist's standpoint 
 there are numerous  reasons  of interest in the
 HT expansions. They are powerful tools for studying numerically 
the properties of statistical models and enable us to test 
 directly, in specific cases, 
 the validity  of the basic assumption of 
universality in the critical phenomena, 
as well as  to assess   the accuracy of the  estimates of the 
universal critical observables 
  obtained   within the renormalization 
group theory by such perturbative approximations as the 
$\epsilon$-expansion  or the fixed-dimension expansion\cite{fwpa}. 

 On the other hand, from a mathematician's standpoint, the 
calculation of long HT expansions
 poses  intriguing challenges of combinatorial, algorithmic
 and analytical nature and, in various cases,
 it  provides generating functions 
  for difficult enumeration problems on a lattice\cite{combi}. 

  We have devoted  this note to  
 a brief discussion   of the use of the Schwinger-Dyson (SD)
equations  in computing HT 
series \cite{bcm86,bccm87,bcm88,bcm90,bc92,bc93,bcg93,bc94,pbmc}
 for the $N$-vector
lattice spin 
models \cite{st68} and in deriving,  as  byproducts, combinatorial results
concerning  the enumeration of 
self-avoiding walks(SAW's)  on  d-dimensional  lattices.

The SD equations have been formulated for several lattice spin 
models\cite{bcm86,bccm87,bcm88,bcm90,bc92,bc93,bcg93,bc94,pbmc}. 
 They are an infinite hierarchy  of coupled linear equations for the spin 
correlation functions. We have shown that they can be effectively used 
as recursion relations to compute HT 
expansions for the correlation functions.
In one form or another and under different denominations,  
these or closely related sets  of equations have long been 
 known in statistical mechanics and in quantum field theory and have been 
 often used in various contexts, in particular  
 recently in lattice field theory (see, for example, 
the very partial list of references\cite{sdlat}). 
It seems however that the SD eqs. have  never been
  systematically used to produce HT series, 
(partial exceptions are Refs.\cite{prec}), 
 and moreover that 
 they are unfamiliar as a tool of the  combinatorial analysis,  
 so that their potential in this context remains largely unexploited. 

  We have  formulated and used  the SDE eqs. for: 
i) the $N$-vector  (also called O($N$) classical 
Heisenberg) model\cite{bcm88,bcm90} on the square and the cubic lattices;
ii)  the XY (or O(2)) model\cite{bcm86,bccm87,bc93,bcg93,bc94} on the square, 
on the triangular  and on the simple cubic lattices; 
iii) the $RP^{N-1}$ model (or generalized Meier-Saupe model)\cite{bc92}
 on the square lattice.

\section{ The SD equations for the $N$-vector model}
The $N$-vector model is defined as follows: to each site $\vec x_i$
of a lattice (to be definite let us consider a cubic lattice)
we associate an $N$-component unit vector $\vec s(\vec x_i)$, called 
{\it spin}, and take as the Hamiltonian 
\begin{equation}
H\{s\}= -\sum_{\langle i,j \rangle} \vec s(\vec x_i)
 \cdot \vec s(\vec x_j) 
\end{equation}
where $\langle i,j \rangle$ indicates that the sum extends to nearest
neighbor sites only.
 Therefore 

\begin{equation}
 Z(\beta) = \int exp( -\beta H\{s\}) \prod_i d\vec s(\vec x_i)
\label{part}
\end{equation}		
is the partition function and $\beta$ the inverse temperature. 
 [Of course, eq.(\ref{part}) and many expressions below have only a formal 
meaning, unless they refer to a finite sublattice $\Lambda$ and the limit, as 
$\Lambda \rightarrow \infty$, is properly defined. 
However, the rigorous reader 
 can easily work out by himself these well known details  that are omitted 
 here  to avoid overburdening the notation.]
If we set 
\begin{equation} 
\phi(C)=\phi(\vec x_1, \vec x_2,..  \vec x_n; \{ b_{i,j} \})=
\prod_{1\le i < j \le n} \big( \vec s(\vec x_i) 
\cdot  \vec s(\vec x_j) \big)^{b_{ij}} , 
\end{equation}
 with $ b_{i,j}\ge 0$,  
the most general correlation function of the model can be written as
\begin{equation}  
\langle \phi(C) \rangle= \frac {1} {Z(\beta)} \int \phi(C)
exp( -\beta H\{s\}) \prod_i d\vec s(\vec x_i)
\end{equation} 
 and therefore it is uniquely identified by the {\it configuration} $C$, 
namely by the
the set of lattice sites $\vec x_1, \vec x_2,..  \vec x_n$,  called 
${\it vertices}$, and by 
the integers $\{ b_{i,j} \}$, called 
${\it number}$  ${\it of}$  ${\it   bonds}$
 between the vertices  $(\vec x_i, \vec x_j)$. This notation  suggests 
a simple graphical representation for any correlation function 
 $\langle \phi(C) \rangle$ on the 
 lattice.
 Using the invariance of the 
 integration measure in the partition function under a site-dependent
 $O(N)$ transformation of the spins, we have shown\cite{bcm88}  
how to obtain a convenient set of SD equations 
 for  the correlation functions of the model:
\FL
\begin{eqnarray}
(N-2+g_1)\langle \phi(C) \rangle= \beta\sum_{\mu=-3}^{3} 
(\langle \phi(C^-_{\mu})\rangle - \langle \phi(C^+_{\mu}) \rangle ) +
 (b_{12}-1)\langle \phi(C_{12,12})\rangle +\sum_{j=3}^{n}b_{1j}
\langle \phi(C^{2j}_{12,1j})\rangle
\label{sdgen}
\end{eqnarray} 

 Here $g_1=\sum_{j=2}^{n} b_{1j}$ denotes 
 the number of bonds connected with the 
vertex $\vec x_1$. 
 The configuration $C^+_{\mu}$ is obtained from the configuration 
$C$ by adding 
the bond $(\vec x_1, \vec x_1 +\vec e_{\mu})$, with 
 $\vec e_{\mu}$ a unit vector in the $\mu$ lattice direction, namely 
\begin{equation}
\phi(C^+_{\mu}) = \vec s(\vec x_1)\cdot\vec s(\vec x_1+\vec e_{\mu}) \phi(C) 
 \end{equation}
 
The configuration $\phi(C^-_{\mu})$ is  obtained from $C$  by changing a bond 
$(\vec x_1, \vec x_2)$ into $(\vec x_1+\vec e_{\mu}, \vec x_2)$,   namely 
\begin{equation}
\phi(C^-_{\mu}) = \frac{\vec s(\vec x_1+\vec e_{\mu}) \cdot \vec s(\vec x_2)}
 {\vec s(\vec x_1)\cdot\vec s(\vec x_2)} \phi(C). 
\end{equation}
 
The configuration $\phi(C_{12,12})$ is  defined by 
\begin{equation} 
\phi(C_{12,12})= \frac{\phi(C)}{(\vec s(\vec x_1)\cdot\vec s(\vec x_2))^2}
\end{equation}
 whenever there are at least two bonds connecting the sites 
$\vec x_1$ and $\vec x_2$ in the configuration $C$.

 Finally, the configuration
 $ \phi(C^{2j}_{12,1j})$ is  obtained from $C$  by removing a bond between 
$(\vec x_1, \vec x_2)$,  one between 
$(\vec x_1, \vec x_j)$ and by adding a new 
 bond between $(\vec x_2, \vec x_j)$, namely 
\begin{equation}
\phi(C^{2j}_{12,1j})= \frac{\vec s(\vec x_2)\cdot\vec s(\vec x_j)}
 {\vec s(\vec x_1)\cdot\vec s(\vec x_2) \vec s(\vec x_1)
\cdot\vec s(\vec x_j)}\phi(C). 
\end{equation}

Since  $C$  can be any configuration, including  the empty configuration 
$C= \emptyset$ associated with the trivial correlation 
$\langle \phi(0) \rangle =1 $,   eqs. (\ref{sdgen}) form
 an infinite hierarchy of linear (inhomogeneous) equations 
 relating all correlation functions.

A simple algorithm solves iteratively this 
system of equations and yields the  
 expansion of any correlation function as a power series of the inverse 
temperature $\beta$. 
The calculation  can be performed on a computer by using exclusively 
 integer arithmetics and therefore is completely
  {\it exact}. Moreover it can be formulated 
 in such a way that  the 
coefficients of the high temperature expansion are explicitly obtained 
as ratios of polynomials in the variable $N$. 

 In order to understand the combinatorial 
relevance of the HT expansion in the 
$N$-vector model,
 it is now sufficient  to recall the  proof\cite{dege} that, in the limit 
 as $N \rightarrow 0$ at fixed $\beta/N$,  the correlation function between 
the spin at the origin  and the spin at  
the site  $\vec x $  is the generating function of the self-avoiding walks 
 connecting the two sites on the lattice, up to a factor $g$ 
(the coordination number of the lattice). In particular,  
the problem of the enumeration of the 
self-avoiding polygons with $n$ sides is solved 
once we determine the $(n-1)-th$ coefficient in the HT expansion 
   of the  correlation function between the spins at 
nearest neighbor sites.
 The most extensive results  available until now for three-dimensional 
 lattices had been obtained by the 
 King's College group\cite{sykes}
 (up to $\beta^{19}$ on the simple cubic lattice, up to $\beta^{15}$ on 
the body-centered cubic lattice and up to  $\beta^{13}$ on the 
face-centered cubic lattice) and date back to more than 25 years ago.
We have recently extended these 
results to order $\beta^{21}$ on both the simple cubic and the body-centered 
cubic lattices 
and hope to reach significantly higher orders in a short time\cite{pbmc}.
 Of course, due to the enormous progress in computers 
in the last quarter of century,  
 the traditional direct  counting techniques employed by the the 
 King's College group  
 would today produce much longer series, 
but the main point of this note is to
 present an interesting alternative  technique rather than an impressive
 computation. 

 Our extended results are:

\[ P_{sc}(\beta)=  24\beta^{3} +264 \beta^{5} + 3312\beta^{7} +
 48240\beta^{9} + 762096\beta^{11} + 12673920\beta^{13} +\]
\[ 218904768\beta^{15} + 3891176352\beta^{17} + 
70742410800\beta^{19} +   1309643747808 \beta^{21}+... \]     	

on the sc lattice and 

\[ P_{bcc}(\beta)=  96\beta^{3} +1776 \beta^{5} + 43776\beta^{7} +
1237920\beta^{9} +   37903776\beta^{11} +  1223681760\beta^{13} +\] \[
 41040797376\beta^{15} +  1416762272736\beta^{17} + 
50027402384640\beta^{19} +   1799035070369856 \beta^{21} +... \]     	

on the bcc lattice.

More results of combinatorial interest emerge, for instance,  by computing, 
in the same limit  $N\rightarrow 0$,
the susceptibility, which is the SAW chain
 generating function etc.. See Ref.\cite{bc98} for recent series extensions.

 For the sake of clarity and brevity,  
  we shall not describe here the iterative 
procedure for the solution of eq.(\ref{sdgen}), but rather only 
the analogous procedure  for  
the slightly simpler case 
in which we set $N=2$, namely for the  so-called $XY$ model\cite{kt}. 
In this case the site variables 
 are 2-component unit vectors 
$ \vec s(\vec x_i) = ( s^{(1)}(\vec x_i),s^{(2)}(\vec x_i)) $
 and  the Hamiltonian  and the correlation functions can be parametrized 
 in a much simpler way:
\begin{equation}
 H\{\theta\} = -\sum_{\langle i,j \rangle} 
\vec s(\vec x_i) \cdot \vec s(\vec x_j) = 
-\sum_{\langle i,j \rangle} cos (\theta_i - \theta_j)
\label{uno}
\end{equation}

 so that 
\begin{equation}
Z = \int_0^{2\pi}  exp( -\beta H\{\theta\}) \prod_i d\theta_i 
\label{due}
\end{equation}

where $\theta_i$ is the angle 
formed by the spin at the site $\vec x_i$ with a given fixed direction.

A complete set of correlation functions for this system can be written
as follows :

\begin{equation}
\langle \varphi({\bf s}) \rangle =
   {1 \over Z} \int_0^{2\pi}  \varphi({\bf s}) exp( -\beta 
H\{\theta\}) \prod_i d\theta_i
\label{cinque}
\end{equation}

\begin{equation}
\varphi({\bf s})= exp( i \sum_{k=1}^N q_k \theta_k), 
\label{sei}
\end{equation}

and $q_k$ is an integer 
number that we call the "charge" at the site $\vec x_k$.
 In  this special case of the $N$-vector model, 
we can put   the correlation functions  into 
a one to one correspondence
with the "configurations of charges" on the lattice, namely 

$ \langle \varphi({\bf s}) \rangle$ 
$\leftrightarrow {\bf s} 
\equiv (\vec x_1,q_1;\vec  x_2,q_2; ...;\vec x_N,q_N) $.

Let us observe that the Hamiltonian has the global symmetry 
 $\theta_i \rightarrow \theta_i + \psi$.
On  two-dimensional lattices, by the Mermin-Wagner theorem, 
the continuous symmetries cannot be broken, 
therefore

\begin{equation}
\langle \varphi({\bf s}) \rangle \not= 0  \iff  \sum_{k=1}^N q_k = 0,
\label{sette}
\end{equation}

at any temperature. 
This means that   nontrivial correlation functions are 
associated only to the globally "neutral" charge configurations. 
 On three-dimensional lattices, this is still true at 
 high temperatures, while the symmetry is broken 
at low temperatures. 

For this model the simplest set of SD eqs.  is the infinite 
hierarchy 

\begin{equation}
 \int_0^{2\pi} \Pi_i d\theta_i {d \over d\theta_k} 
 [\varphi({\bf s}) exp( -\beta H\{\theta\}) ] = 0 
\label{otto}
\end{equation}

 for all possible  $\varphi({\bf s})$ and taking  $\vec x_k \in {\bf s}$.
These identities are an obvious consequence of the $2\pi$ periodicity
of the quantity in square parentheses. More explicitly we have 

\begin{equation}
\langle \varphi({\bf s}) \rangle = 
 {\beta \over  2q_k}
 \sum_{\mu} \Big( \langle 
exp[- i(\theta_k- \theta_{k+\mu})]\varphi({\bf s})\rangle - 
\langle exp[ i(\theta_k- \theta_{k+\mu})]\varphi({\bf s}) \rangle \Big ) 
\label{nove}
\end{equation}

where the sum extends to all of the nearest neighbor sites of $\vec x_k $. 
Introducing the notation
 ${\bf s^+_{\mu}} ({\bf s^-_{\mu}})$
for the lattice charge configuration obtained increasing 
(decreasing) by 1 the charge at 
the site ${ \vec x_k}$ and decreasing (increasing) by 1 the charge at 
the site $ \vec x_{k}+\vec e_\mu$,
 we can write  eqs. (\ref{nove}) more compactly
as follows:

\begin{equation}
\langle \varphi({\bf s}) \rangle = 
{\beta \over 2q_k} \sum_{\mu} \Big(\langle \varphi({\bf s^-_{\mu}}) \rangle - 
\langle \varphi({\bf s^+_{\mu}}) \rangle \Big ).
\label{dieci}
\end{equation}

For example, on the square lattice let us consider  the charge 
configuration ${\bf s}$ = $ (0,0,1; 1,0,-1)$, so that 
$\langle \varphi({\bf s}) \rangle$ is the elementary 
nearest neighbor correlation function  
$\langle exp[ i (\theta_{00} - \theta_{10})] \rangle$. Let us also assume
that the derivative $ d \over d \theta_k$ acts on the first site of each 
lattice charge configuration, namely take $k=1$, then ${\bf s^+_{\mu}}$ and
 ${\bf s^-_{\mu}}$ represent the following four configurations

\FL
\begin{eqnarray}
 {\bf s^+_{\mu}} \equiv \big\{ (0,0,2;1,0,-2), 
(0,0,2;1,0,-1;0,1,-1),
(0,0,2;1,0,-1;-1,0,-1),\nonumber\\ 
 (0,0,2;1,0,-1;0,-1,-1)\big\}
\label{undici}
\end{eqnarray}

\FL
\begin{eqnarray}
 {\bf s^-_{\mu}} \equiv \big\{ \emptyset, (0,1,1;1,0,-1), 
(-1,0,1;1,0,-1), (0,-1,1;1,0,-1)\big\}
\label{dodici}
\end{eqnarray}

It is easy now to understand intuitively how  to produce a 
HT expansion: we have only to
 notice that in the previous example the correlation function 
$\langle exp[ i (\theta_{00} - \theta_{10})] \rangle$ on the lhs 
 of the SD eq. has a HT 
expansion  beginning 
at order $\beta$, while on the rhs there 
is a linear combination of correlation 
functions whose HT expansions   begin
 at order $\beta^2$, except for $\langle \varphi({\bf 0}) \rangle = 1$. 
 Therefore we obtain

\begin{equation}
\langle exp[ i (\theta_{00} - \theta_{10})] \rangle =  
{\beta \over 2} (1+ O(\beta^2)).
\label{tredici}
\end{equation}

This remark immediately yields the first term of 
the HT expansion for the nearest 
neighbor correlation 
function. 
The successive term, (which is 
actually $O(\beta^3)$), is obtained by the same mechanism  using the SD eqs. 
 for the correlation functions  $\langle \varphi({\bf s^-_{\mu}}) \rangle$ 
 and
$\langle \varphi({\bf s^+_{\mu}}) \rangle $ in 
the  $O(\beta^2)$  terms  on the rhs and so on.
 Once this simple mechanism has been made clear, it is convenient to 
 formulate
the  procedure in a way which also suggests how to make 
it completely automatic and therefore ready to be  translated  
into a computer code.

\noindent To this purpose let us first define a linear space 
of states as follows:
 
Consider the set of all possible distinct neutral charge configurations
on the lattice. Two such charge configurations 
$ {\bf s} \equiv (\vec x_1,q_1;\vec  x_2,q_2; ...;\vec x_N,q_N) $ 
and $ {\bf s'} \equiv (\vec x'_1,q'_1; \vec x'_2,q'_2; ...;\vec x'_M,q'_M) $
 are considered distinct if  they cannot be 
identified by a permutation of labels and/or by a lattice roto-translation 
and/or by changing the sign of all charges. To each charge configuration 
${\bf s}$,  we associate a state vector 
represented by a ket $\vert {\bf s} \rangle$;
 to the empty lattice ${\bf s= \emptyset}$, we associate the ket
$\vert {\bf 0} \rangle$.
We define a scalar product as follows: $\langle {\bf s} \vert {\bf s'} 
\rangle = 1$ if ${\bf s} = {\bf s'} $  
and $\langle {\bf s} \vert {\bf s'} 
\rangle = 0$ otherwise.
The set of all possible states is therefore an orthonormal basis of an 
infinite-dimensional linear vector space on which we can define a linear 
operator $K$ as follows:

\begin{equation}
K \vert {\bf 0} \rangle= 0
\label{quattordici}
\end{equation}

\begin{equation}
K \vert {\bf s} \rangle = 
{1 \over 2q_1} \sum_{\mu} 
\Big( \vert {\bf s^-_{\mu}} \rangle -  \vert {\bf s^+_{\mu}} \rangle \Big)
\label{quindici}
\end{equation}

here $q_1$ is the charge at the site $\vec x_1$.

Now consider the identity operator 

\begin{equation}
1 = (1 - \beta K) (1 - \beta K)^{-1}
\label{sedici}
\end{equation}

and take its matrix element between the states $\langle {\bf 0} \vert$ and 
$\vert {\bf s} \rangle$, namely: 

\begin{equation}
\langle {\bf 0} \vert (1 - \beta K) (1 - \beta K)^{-1} 
\vert {\bf s} \rangle = 0 .
\label{diciassette}
\end{equation}

 From (\ref{diciassette}) it follows :

\begin{equation}
\langle {\bf 0} \vert (1 - \beta K)^{-1} \vert {\bf s} \rangle = \beta
\langle {\bf 0} \vert (1 - \beta K)^{-1} K \vert {\bf s} \rangle .
\label{diciotto}
\end{equation}

Using the definition of $K$ (\ref{quattordici}),(\ref{quindici}) 
we have finally:

\begin{equation}
\langle {\bf 0} \vert (1 - \beta K)^{-1} \vert {\bf s} \rangle  =
 {\beta \over 2q_1} \sum_{\mu} \langle {\bf 0} \vert (1 - \beta K)^{-1} 
\Big( \vert {\bf s^-_{\mu}} \rangle  - 
 \vert {\bf s^+_{\mu}} \rangle \Big) .
\label{diciannove}
\end{equation}

We conclude  that 
$\langle {\bf 0} \vert (1 - \beta K)^{-1} \vert {\bf s} \rangle$ 
 satisfies the same SD eq. as $\langle \varphi( {\bf s}) \rangle$ and 
therefore  

\begin{equation}
\langle \varphi ({\bf s}) \rangle  = 
\langle {\bf 0} \vert (1 - \beta K)^{-1} \vert {\bf s} \rangle.
\label{venti}
\end{equation}

Now the solution of the SD eqs. can simply be obtained by a power series 
expansion of the resolvent operator $(1- \beta K)^{-1}$. We have

\begin{equation}
\langle \varphi( {\bf s}) \rangle = \sum_n \beta^n \langle {\bf 0} 
\vert  K^n \vert {\bf s} \rangle. 
\label{ventuno}
\end{equation}

This is precisely the HT expansion of 
$\langle \varphi( {\bf s}) \rangle$. The n-th order coefficient 

\begin{equation}
c_n =\langle {\bf 0} \vert  K^n \vert {\bf s} \rangle
 = \sum_{ s_1,s_2,..s_n}\langle {\bf 0} \vert  K \vert {\bf s_1} \rangle
\langle {\bf s_1} \vert  K \vert {\bf s_2} \rangle...
\langle {\bf s_{n-1}} \vert  K \vert {\bf s} \rangle  
\label{ventidue}
\end{equation}

is a particular matrix element of the n-th power of the matrix 
$\langle {\bf s }\vert  K \vert {\bf s'} \rangle$.

 The structure of the matrix $K$ is quite simple: it is a (very) sparse 
matrix with elements 

\begin{equation}
\langle {\bf s} \vert  K \vert {\bf s'} \rangle
 = {\pm 1 \over 2q_1} 
\label{ventitre}
\end{equation}

 if $ {\bf s = s'^{\pm}_{\mu}}$ and 0 otherwise.

Therefore in every column there appear at most $2g$  nonvanishing elements 
($g$ being the 
coordination number of the lattice).

In order to calculate a HT series 
up to some maximal order $\beta^M$, it is 
sufficient to compute the matrix $K$ on a finite truncated basis.
We can then implement the following iterative procedure for computing the HTE
 of the correlation function $\langle \varphi ({\bf s}) \rangle$ up to the 
order M.  

In order to be  definite we shall continue to consider 
 the case of a square lattice.

First iteration: 

consider the state $\vert {\bf s} \rangle  $ and call it
 state $\vert {\bf 1} \rangle $. Compute $ {\bf K} \vert {\bf 1} \rangle $.
This produces the 8 new states 
$ {\vert \bf s_{\mu}^{\pm}\rangle } = (  \vert {\bf 2} 
\rangle,  \vert {\bf 3} \rangle,.. , \vert {\bf 9} \rangle )$.
 Memorize the matrix elements of column 1, namely:

 $K_{21}= \langle {\bf 2} \vert K \vert {\bf 1} \rangle$,
$ K_{31}= \langle {\bf 3} \vert K \vert {\bf 1} \rangle$,..., 
$ K_{91}= \langle {\bf 9} \vert K \vert {\bf 1} \rangle $.

Clearly these are  the only nonvanishing elements of the first column of 
the matrix ${\bf K}$.

(Remember that the operator ${\bf K}$ acts only on the "first" charge
 of each state, which can be arbitrarily chosen.)

Second iteration: 

Act with ${\bf K}$ on the new states 
$  \vert {\bf 2} 
\rangle,  \vert {\bf 3} \rangle,.. , \vert {\bf 9} \rangle $. Then $8^2$
 new configurations are produced:  $  \vert {\bf 10} \rangle, 
 \vert {\bf 11} \rangle,.. , \vert {\bf 73} \rangle $. Memorize now
the nonvanishing elements of columns 2,3,..9 namely 
$K_{10,2}, K_{11,2},..., K_{73,9}$.

 Continue until M iterations are completed.

This is a correct, but definitely  naive and inefficient procedure: 
the growth of the truncated basis on which we compute 
the elements of the matrix $K$ is very fast. 
It appears that, in order to perform the (M+1)-th iteration, we 
should store  $8^M$ states
 ( or $(2g)^M$ for a  general lattice with coordination number $g$). 
Therefore $M \simeq 10$, or even less
than that, would seem to be an insuperable 
upper limit for this kind of computation.
However, only a small subset of the states produced in the 
iteration are  distinct with respect to the natural equivalence relation 
mentioned above. 
We must obviously exploit   this fact 
 by always identifying equivalent states, 
namely by working with equivalence classes 
 of states.  This  reduces 
 drastically the growth of 
the truncated basis. For example, on the square lattice, the growth is 
reduced from $8^M$ to some $(\simeq 3)^M$. Another very useful observation is 
that, 
after the $[M/2]$-th iteration, an increasing fraction of the new 
states produced do not actually contribute to the matrix elements of 
$\langle {\bf 0} \vert  K^r \vert {\bf s} \rangle$, provided that $r \le M$. 
Therefore these states are irrelevant for the calculation 
and can be simply discarded. This entails a 
further reduction of the dimension of the truncated basis to some more 
manageable $(\simeq 2)^M$. These simple remarks show why much higher 
orders than $M \simeq 10$ can be attained.

We have sketched 
  the simplest recursive approach to the  
SD eqs. (\ref{sdgen}). The    procedure
 needed for the general $N$-vector  model is only slightly more intricate
 and the interested reader can find all details  in Ref.\cite{bcm88}.

\section{Conclusions}

In conclusion, we stress that conceptual simplicity
 is the main advantage of the SD method. 
 It enables us to bypass the usual elaborate  graphical formulation 
of the  HT expansion 
 procedures  in statistical mechanics, 
 namely  the construction and selection of   the relevant graphs, 
 the computation of  the graph embedding constants 
in the lattice etc., 
 as well as  any other laborious combinatorial inputs. 
 Indeed  all correlation functions of  the system
under study are  produced directly by an elementary recursion.
 Moreover  the method is very flexible 
and widely  applicable.
 This simplicity also entails the limitations 
of the  technique: after iterating sufficiently many times
 the recursion relations, a plethora 
of contributions is generated  and, for each given model,
 it takes 
 some ingenuity to classify them and 
 to organize efficiently the computation
 by managing carefully 
 and economically  the storage resources of the computer.
 Moreover, due to presence of a summation over $\mu$ in eqs. (\ref{sdgen})
 and (\ref{dieci}), the number of contributions at a given iteration
 increases rapidly with the dimensionality of the lattice.
 In our first tests of the method, the maximal 
order of expansion was limited only by the size of the 
computer memory, (which, only  a few years ago,  was usually 
 two orders of magnitude 
smaller than it is today in a common pc), 
whereas the  required CPU times were  always 
very modest.  For this reason, after performing a few computations,  
we had set aside the recursion methods and 
   turned to  the traditional 
linked-cluster-expansion techniques\cite{bc98}.  
 We believe however, that   the Schwinger-Dyson technique will be 
increasingly useful and easy to use  as larger 
 mass memories  become available in computers. 
 Certainly, the very impressive results obtained by 
algebraic techniques\cite{cge}
 in low-dimensional models with discrete site-variables
(for instance SAW and Ising models)  are unlikely to be improved 
or even to be reproduced by these procedures. 
The recursion approach,  however, 
remains superior for models with continuous variables, like the XY model,
 where the algebraic methods become inefficient.
Therefore, we believe that the Schwinger-Dyson method
 deserves further exploration in a wider context, 
both from the formal and from the algorithmic point of view.

\end{document}